\documentclass{emulateapj}
\usepackage{graphicx}
\usepackage{amsmath}
\usepackage{units}
\usepackage{natbib}
\usepackage{color}
\usepackage{wasysym}

\newcommand{\figext}[1]{figures/#1.pdf}

\shorttitle{Sky maps of GRB 170817A}

\begin{document}
\title{Radio sky maps of the GRB 170817A afterglow from simulations}

\author{Jonathan Zrake\altaffilmark{1}, Xiaoyi Xie\altaffilmark{2}, Andrew MacFadyen\altaffilmark{2}}
\affil{$^1$Physics Department and Columbia Astrophysics Laboratory,
Columbia University, 538 West 120th St, New York, NY 10027, USA}
\affil{$^2$Center for Cosmology and Particle Physics, Physics Department,
New York University, 726 Broadway, New York, NY 10003, USA}










\begin{abstract}
We present synthetic radio images of the GRB170817A afterglow, computed from moving-mesh hydrodynamic simulations of binary neutron star merger outflows. Having expanded for nearly a year, the merger remnant is expected to subtend $\sim5$ milli-arcseconds on the sky, potentially resolvable by very long baseline radio imaging techniques. Any observations revealing the radio centroid to be offset from the line-of-site to the merger would be the smoking gun of a jetted outflow. However, our results indicate that a measurement of the centroid position alone cannot independently determine whether that jet escaped successfully from the merger debris cloud, or was ``choked,'' yielding a quasi-spherical explosion. We find that in both scenarios, the centroid exhibits superluminal proper motion away from the merger site at roughly 4 -- 10 micro-arcseconds per day for at least the first 300 days. We argue that a successful strategy for differentiating among the explosion models will need to include multiple observations over the coming months -- years. In particular, we find the time at which the centroid attains its maximum offset, and begins heading back toward the merger site, is considerably later if the jet was choked. Detecting a reversal of the centroid trajectory earlier than 600 days would uniquely identify a successful jet. Our results indicate the source might be resolved using VLBI radio observing techniques with $\sim\unit[1]{mas}$ resolution starting at roughly 400 days post-merger, and that the the angular extent of a successful jet is significantly smaller than that of a choked jet (4.5 versus 7 mas respectively).

\end{abstract}

\maketitle

\section{Introduction}
\label{sec:introduction}
The short gamma-ray burst (sGRB) GRB170817A is the first discovered electromagnetic counterpart to a gravitational wave (GW) merger event. Its prompt emission was detected shortly following the ``chirp'' of GW170817 \citep{lvc170817, em170817}, caused by the inspiral and coalescence of a binary neutron star (BNS) system in NGC4993, roughly $\unit[40]{Mpc}$ away \citep{Coulter2017, Soares-Santos2017}. Among short bursts with host galaxy identifications, GRB170817A is the nearest to the earth \citep[e.g.][]{Fong2017}, and has the lowest isotropic-equivalent gamma-ray energy, $\sim \unit[10^{46} - 10^{47}]{erg}$ \citep{Goldstein2017, Savchenko2017}. However, X-ray through radio frequency follow-up observations \citep{alex17, Haggard17, hal17, Kasliwal17, marg17, Troja17, Margutti2018, mool18} indicate that the explosion was intrinsically far more energetic, likely exceeding $\sim \unit[10^{51}]{erg}$. The afterglow emission is also exceptional in that it rose steadily for $\sim \unit[160]{days}$, showing a turnover only after $\sim \unit[200]{days}$ \citep{alexander18, dobie18, Nynka2018}.

A key question raised by these observations is whether GRB170817A is a classical sGBR seen significantly off-axis, or represents a new class of astrophysical transient \citep{Kasliwal17, np18, Hotokezaka2018}. Numerical modeling \citep{laz17, Gottlieb2017, Bromberg2018, Xie2018} has demonstrated that the broadband afterglow light curves of GRB170817A can be produced by either of two distinct explosion scenarios, differing primarily by their degree of anisotropy. These are commonly referred to as the ``successful jet'', which could be seen by distant, on-axis observers as a cosmological sGRB, and the ``choked jet'' scenario, which would only be observable nearby. Since multi-band light curves do not conclusively distinguish between these scenarios, we are motivated to explore other observational signatures capable of discerning the outflow geometry. Due to the fortuitous proximity of GW170817, the GRB170817A remnant should currently have an angular size of $\sim 5$ mas on the sky and may thus be spatially resolvable using very long baseline radio observing techniques. \footnote{Since the original submission of this manuscript, VLBI radio observations by \cite{Mooley2018} have been published. The revised version of this Letter contains comparisons of our results with those observations, however all of our results were obtained without any knowledge of them.}

Predictions of the radio image morphology have been computed from adaptive mesh hydrodynamic simulations by \cite{Granot2018} and \cite{Nakar2018}. Morphological and linear polarization features based on parameterized jet models were also reported in \cite{Gill2018}.

To enhance the ability of potential radio imaging to distinguish among the as yet degenerate explosion geometries, we present in this Letter radio synchrotron sky maps computed from our previously published \citep{Xie2018} moving-mesh hydrodynamic simulations. We focus our analysis on the time evolution of the radio image morphology. Linear polarization maps will be reported in the near future.

\begin{figure*}
  \includegraphics{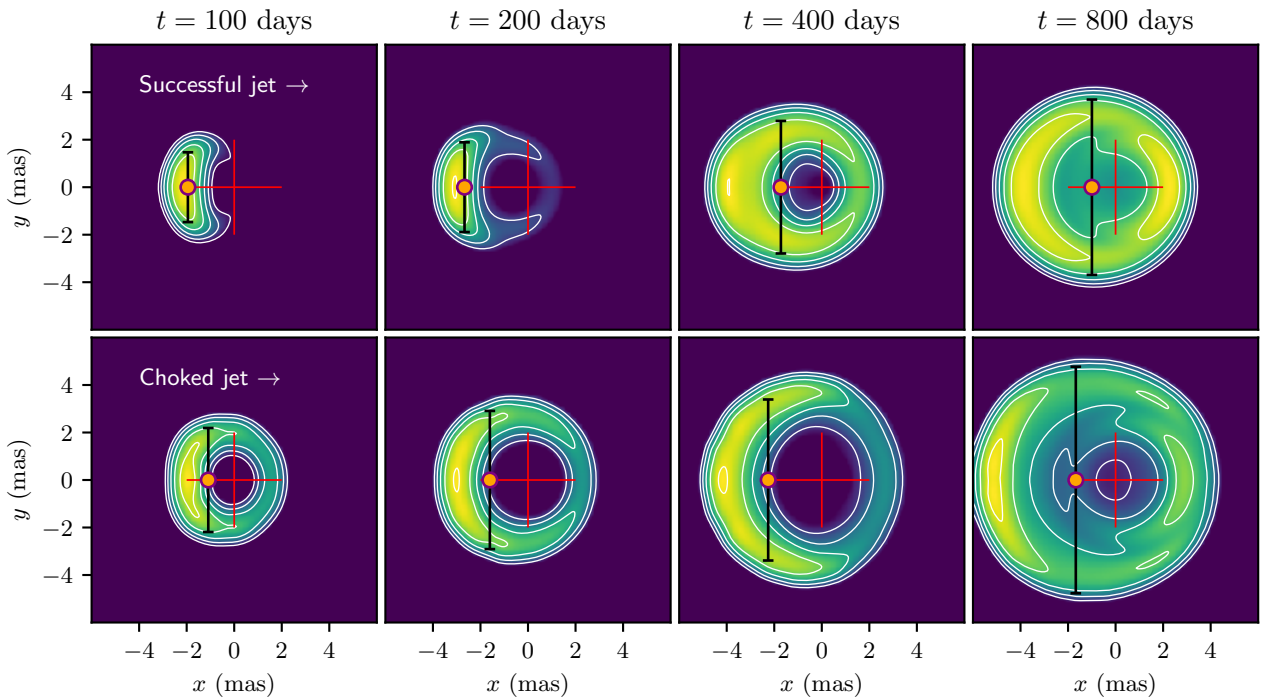}
  \caption{Images of the intensity distribution, $I_\nu(x, y)$ for both explosion models seen at $\theta_{\rm obs} = 20^\circ$. $I_\nu(x, y)$ is represented logarithmically, increasing in value over two decades from $\log_{10} I_{\nu, \rm max} - 2$ (dark blue) to $\log_{10} I_{\nu, \rm max}$ (yellow). The jet axis is oriented horizontally, with the approaching side on the left and the receding side on the right. The red plus sign marks the merger site, and is $\unit[2]{mas}$ in size. The filled orange circles mark the centroid $x_c$ of the intensity distribution. The vertical black bars are positioned horizontally at $x_{\rm max}$, where the longitudinal intensity distribution $I_{\nu, \rm avg}(x)$ peaks, and their height is $\delta y$, the FWHM of $I_\nu (x_{\rm max}, y)$. $x_c$ and $\delta y$ are computed at $\unit[5]{GHz}$, while the logarithmic image morphology is frequency-independent for any frequency on the same power-law segment of the synchrotron spectrum.}
  \label{fig:image-comparison-20}
\end{figure*}

\begin{figure*}
  \includegraphics{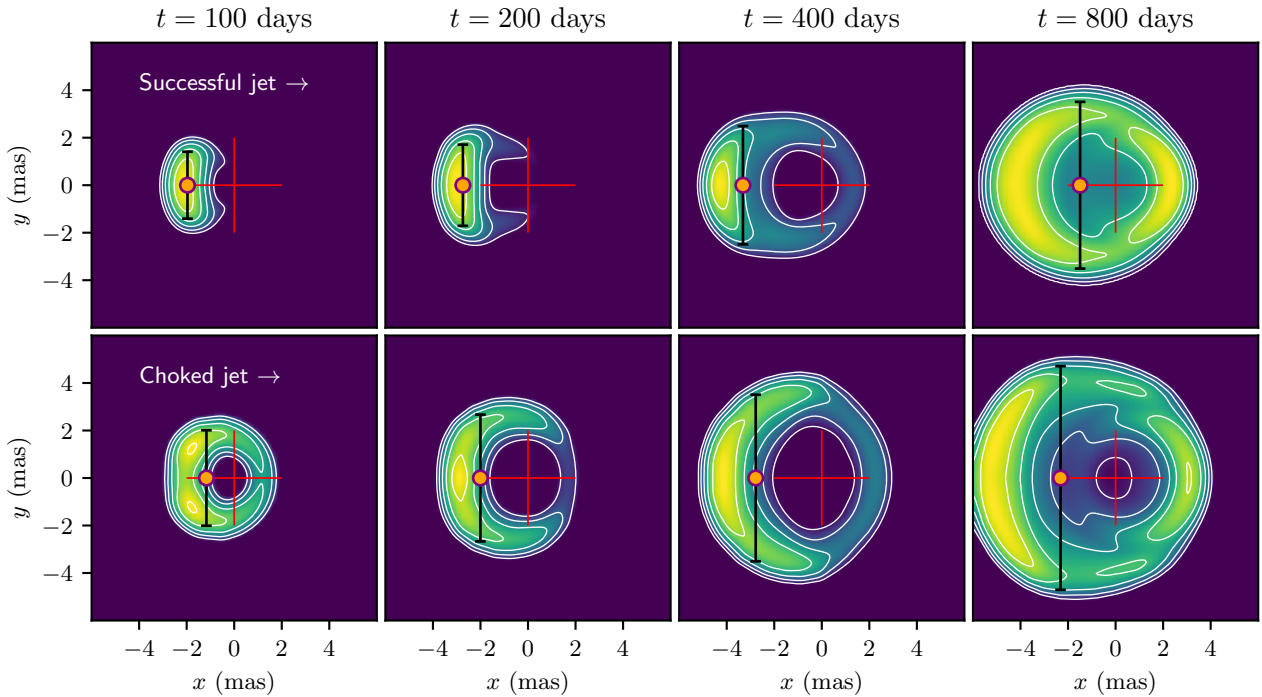}
  \caption{Images of the intensity distribution, $I_\nu(x, y)$ for both explosion models seen at $\theta_{\rm obs} = 30^\circ$. $I_\nu(x, y)$ is represented logarithmically, increasing in value over two decades from $\log_{10} I_{\nu, \rm max} - 2$ (dark blue) to $\log_{10} I_{\nu, \rm max}$ (yellow). The jet axis is oriented horizontally, with the approaching side on the left and the receding side on the right. The red plus sign marks the merger site, and is $\unit[2]{mas}$ in size. The filled orange circles mark the centroid $x_c$ of the intensity distribution. The vertical black bars are positioned horizontally at $x_{\rm max}$, where the longitudinal intensity distribution $I_{\nu, \rm avg}(x)$ peaks, and their height is $\delta y$, the FWHM of $I_\nu (x_{\rm max}, y)$. $x_c$ and $\delta y$ are computed at $\unit[5]{GHz}$, while the logarithmic image morphology is frequency-independent for any frequency on the same power-law segment of the synchrotron spectrum.}
  \label{fig:image-comparison-30}
\end{figure*}

\section{Numerical methods}
\label{sec:methods}

\subsection{Summary of simulations}
The sky maps presented in this Letter are based on our previously published numerical simulations \citep{Margutti2018, Xie2018}, which we briefly summarize here. We utilize the moving-mesh relativistic hydrodynamics code {\tt JET} \citep{Duffell2013} to perform 2D, axi-symmetric simulations of jets launched into a cloud of expanding neutron star merger ejecta. 
The ejecta cloud mass is based on observations of kilonova emission indicating that $\sim 10^{-2} \ M_\odot$ of material was ejected during the merger and its aftermath. We employ a cloud structure motivated by numerical relativity simulations \citep[see][for details of the initial conditions]{Xie2018}. In order to follow the fluid evolution from relativistic to non-relativistic conditions, we have employed an equation of state (EOS) that well approximates the exact EOS for a perfect single-component gas \citep{ryu2006}.

We consider two distinct models for the central engine: (i) a narrow engine ($\theta_{\rm jet} = 0.1$), producing a well-collimated outflow that breaks out of the debris cloud with ultra-relativistic velocity ($\Gamma \sim 100$) and quasi-Gaussian angular structure $\Gamma(\theta)$, and (ii) a wide engine ($\theta_{\rm jet} = 0.35$) leading to a slower ($\Gamma \lesssim 10$) quasi-spherical explosion. These two classes of outflow are commonly referred to as the ``successful structured jet" and ``choked jet" respectively. Both engine models have a jet luminosity $L_{\rm jet} = \unit[2.6 \times 10^{50}]{erg/s}$ and duration $t_{\rm jet} = \unit[2]{s}$. We found previously \citep{Xie2018} that synchrotron light curves from both scenarios, computed with standard radiation modeling \citep{Sari1998}, are broadly consistent with the available observations of the GRB170817A afterglow emission.

\subsection{Calculation of the radio sky images}
Our synthetic radio images are computed directly from hydrodynamic simulation data using a standard GRB afterglow modeling procedure. This analysis is straightforward for optically thin sources \citep[for details see e.g.][note that we neglect synchrotron self-absorption]{Sari1999, DeColle2012}. The specific intensity in a direction $\hat n$ is given by
\begin{equation}
    I_\nu(\hat n, t_{\rm obs}) = \int j_\nu(\hat n, t_{\rm ret}(\mathbf r), \mathbf r) \ d\ell \, ,
\end{equation}
where the integration is along the line of sight through the source, and the integrand is the observer frame synchrotron emissivity, evaluated at the source position $\mathbf r$ and retarded time $t_{\rm ret}(\mathbf r)$. Our numerical procedure is to iterate through the hydrodynamic simulation cells at a discrete set of ($\sim 1000$, logarithmically spaced) lab frame times $\{t^n_{\rm lab}\}$, computing the contribution of each to a three-dimensional histogram of discrete intensity values $\mathcal{I}^m_{i,j}$. The index $m$ corresponds to observer time levels, binned in 10 day intervals, and $i,j$ are the image pixel indexes. Note that we suppress the subscript $\nu$ in the discretized intensity. The histogram is populated according to
\begin{equation}
\label{eqn:discrete-intensity}
    \mathcal{I}^m_{i, j} = \frac{1}{\Delta S_{i,j} \Delta t^m_{\rm obs}} \ \sum_{n,q} j^n_q(\hat n) \Delta V_q \Delta t_{\rm lab}^n \, ,
\end{equation}
where the sum includes those computational cells lying within the image pixel $i,j$, and whose radiation is received during the time interval $\Delta t^m_{\rm obs}$. In Equation \ref{eqn:discrete-intensity}, $\Delta V_q$ is the spatial volume of simulation cell $q$, $\Delta S_{i,j} = |\Delta \mathbf S_{i,j}|$ is the area (in $\unit[]{cm^2}$) of its confining image pixel $(i,j)$ projected to the source location, and $\hat n$ is the unit vector orthogonal to $\Delta \mathbf S_{i,j}$ (and thus pointing from the source to the observer). Because our simulations are axi-symmetric, the computational cells iterated over in Equation \ref{eqn:discrete-intensity} are obtained by sub-dividing the 2D $r-\theta$ simulation volumes into 50 evenly spaced azimuthal cells. This procedure is redundant when computing images or lightcurves for on-axis observers, but is necessary when considering the case of off-axis observers.

\begin{figure*}
  \includegraphics{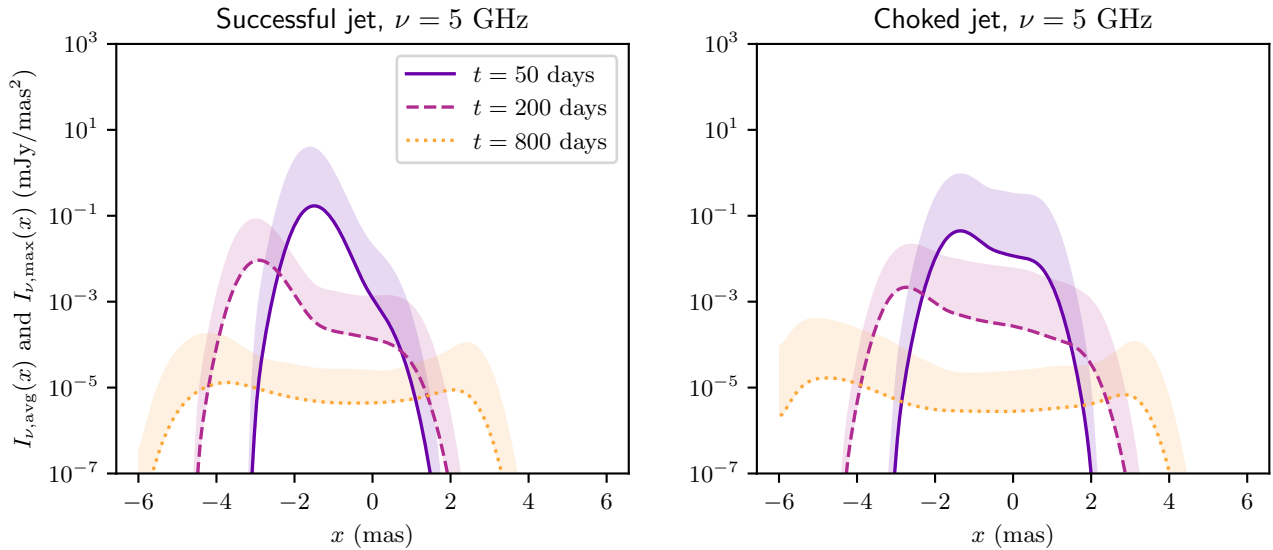}
  \caption{Longitudinal intensity distributions at various observer times for the successful (left) and choked (right) jets seen $30^\circ$ off-axis. The shaded regions are bounded below by the vertically averaged intensity $I_{\nu, \rm avg}(x)$, and above by the maximum $I_{\nu, \rm max}(x)$.}
  \label{fig:vertically-averaged-intensity}
\end{figure*}

The cell emissivity $j^n_q(\hat n)$ is obtained by transforming the comoving emissivity $j'_{\nu'}$ to the lab frame. $j'_{\nu'}$ is isotropic, and describes synchrotron emission from a single power-law distribution of electron energies (having index $p = 2.15$) extending between a minimum synchrotron frequency $\nu'_m$ and a cooling frequency $\nu'_c$ \cite[e.g.][]{Sari1998}. Radio through X-ray observations from 220 days \citep{Alexander2018} showed no sign of a cooling break in the synchrotron spectrum, so radio frequency images based on a single power-law electron distribution are expected to remain valid through late evolution phases. In characterizing the synchrotron emissivity, we adopt nominal parameters $\epsilon_B = 10^{-2}$ and $\epsilon_e = 0.1$ which were successful in \cite{Xie2018} at fitting the synchrotron afterglow lightcurve.

We present images in terms of flux density per solid angle (units of $\unit[]{Jy/mas^2}$) assuming a source distance of $\unit[40]{Mpc}$. For each time bin $m$, $\mathcal{I}^m_{i,j}$ is convolved with an isotropic Gaussian kernel 5 pixels ($\unit[300]{\mu as}$) across. These smoothed histograms are referred to as images, sky maps, or intensity distributions, and are denoted by $I_\nu(x, y)$, where $x$ and $y$ are measured in milliarcseconds. The image coordinate system is centered at the merger site, and oriented so that $-x$ increases along the approaching jet's projection on the sky (the counter-jet is on the right-hand-side of the images where $x > 0$). Given the relatively close proximity of the source, we do not account for cosmological redshift factors.

Our sky maps are created for a sampling of jet observer angles $\theta_{\rm obs}$ (angle between the jet axis and the line-of-sight), between $20^\circ$ and $45^\circ$. This range covers the best-fit observer angles determined by a number of independent groups \citep{Mooley2018, Troja2018, Hotokezaka2018a}. We adopt a nominal observer frequency of 5 GHz.

\section{Results}
\label{sec:results}

Radio sky maps for observer angles $20^\circ$ and $30^\circ$ are shown in Figures \ref{fig:image-comparison-20} and \ref{fig:image-comparison-30}. These are images of the normalized intensity distribution,
\begin{equation}
  \bar I_\nu(x, y) \equiv \frac{I_\nu(x, y)}{I_{\nu, \rm max}} \, ,
\end{equation}
where $I_{\nu, \rm max}$ is the intensity of the brightest pixel in the image. Note that $\bar I_\nu(x, y)$ conveys only the source morphology, not the brightness of individual features from one image to another. In Figure \ref{fig:vertically-averaged-intensity} we show $y$-averaged intensity distributions,
\begin{equation}
  I_{\nu, \rm avg}(x) = \frac{1}{\Delta y} \int I(x,y) \ dy \, ,
\end{equation}
indicating the relative brightness of morphological features and between temporal slices. Also note that the images shown in Figures \ref{fig:image-comparison-20} and \ref{fig:image-comparison-30} depict the logarithm of intensity, and so the image morphology is independent of frequency $\nu$, to the extent that $\nu'$ lies on the same power-law segment of the spectral emissivity (extending from $\nu_m'$ to $\nu_c'$) for all simulation cells. In practice we find that a small number of cells, typically those immediately behind the external shock, contain sufficiently energetic electrons that their minimum synchrotron frequency $\nu_m$ marginally exceeds $\unit[5]{GHz}$.

The images are generally characterized by a crescent-moon shape opening toward the merger site (located at the origin). This feature is synchrotron emission from the approaching relativistic shell, in its decelerating but pre-Sedov phase. As the shell decelerates, the Doppler beaming of its emission lessens, and more of its surface comes into view causing the crescent to grow larger. Emission from the receding shell comes into view at later times, (e.g. between 400 and 800 days in the successful jet model seen at $\theta_{\rm obs} = 30^\circ$), forming a double-ring pattern on the sky.

The successful and choked jet scenarios both generate anisotropic explosions, and yield correspondingly anisotropic sky images. The degree of anisotropy may be characterized by the angular separation between the flux centroid,
\begin{equation}
  x_c \equiv \frac{\int x \ I_\nu(x, y) \ dA}{\int I_\nu(x, y) \ dA} \, ,
\end{equation}
and the line-of-sight to the merger. If the explosion were spherically symmetric, or if an axi-symmetric explosion were viewed on-axis, then $x_c$ would remain at the origin. In other words, observations revealing the flux centroid to be offset from the merger location would rule out both a spherical explosion, and an axi-symmetric explosion viewed on-axis.

The flux centroid $x_c$ is depicted as an orange dot in Figures \ref{fig:image-comparison-20} and \ref{fig:image-comparison-30}. In both models, and at each of the observer angles $\theta_{\rm obs} = 20^\circ$, $30^\circ$, and $45^\circ$, $x_c$ is found to lie at a distance of between 1 and 3.5 mas from the merger site at times between 200 and 400 days. Thus, radio observations with resolution better than $\sim \unit[1]{mas}$ should be able to firmly distinguish between spherical and aspherical explosion scenarios.

\begin{figure*}
  \includegraphics{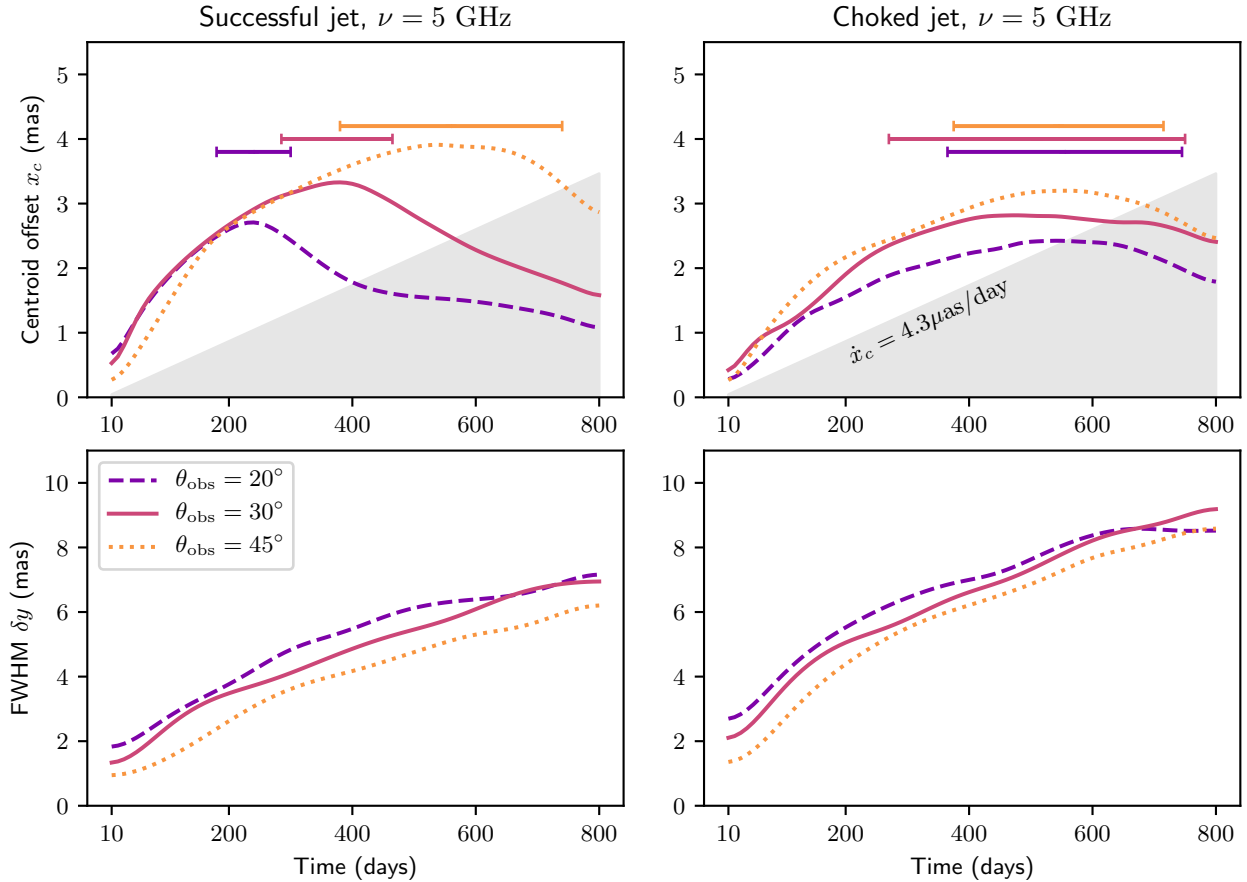}
  \caption{\emph{Top panel}: Evolution of the centroid $x_c$ (filled orange circles in Figures \ref{fig:image-comparison-20} and \ref{fig:image-comparison-30}). The shaded region indicates the angular offset within which the centroid's proper motion is subluminal. The horizontal bars bracket $t_{\rm max}$ and indicate the time interval during which the flux centroid $x_c$ is within $\unit[100]{\mu as}$ of its maximum value. \emph{Bottom panel}: Evolution of the transverse extent $\delta y$ (black bars in Figures \ref{fig:image-comparison-20} and \ref{fig:image-comparison-30}).}
  \label{fig:centroid-extent-comparison}
\end{figure*}

Unfortunately, differentiating between the successful and choked jet models may be more challenging. We show in the top row of Figure \ref{fig:centroid-extent-comparison} the temporal evolution of $x_c$ for both models, and at several observer angles. For reference we also plot the region (shown as a gray wedge) above which superluminal proper motion would be observed, that is where $\dot x_c > c / \unit[40]{Mpc} \simeq \unit[4.3]{\mu as/day}$. We find that in each case, $x_c$ exhibits super-luminal proper motion of at least $\unit[10]{\mu as/day}$ for the first $\sim \unit[200]{days}$, with the exception of the choked jet model at small viewing angle $\theta_{\rm obs} = 20^\circ$, which slows to $\unit[4]{\mu as/day}$ at roughly 200 days.

We have made an effort to identify other features that might aid in differentiating between the successful and choked jet scenarios. The first is the time, $t_{\rm max}$, at which the flux centroid reaches its maximum distance from the merger site and begins moving back toward it. Such change in the direction of centroid motion is inevitable as the explosion enters the Sedov phase and becomes fully spherical. As shown in Figure \ref{fig:centroid-extent-comparison}, if the viewing angle is $20^\circ$, then observations of a structured jet between 200 and 400 days will reveal the flux centroid to be heading back toward the merger site at a rate of roughly $\unit[10]{\mu as/day}$. If the viewing angle is $30^\circ$, then the structured jet scenario predicts that $t_{\rm max} \simeq \unit[400]{days}$ and that $x_c$ remains within $\unit[100]{\mu as}$ of its maximum value for roughly 200 days. In contrast, for a choked jet seen at $\theta_{\rm obs} = 30^\circ$ $x_c$ should exhibit no detectable change (stalling at 2 -- 3 mas) throughout the time interval 200 -- 800 days.

Another feature that may help differentiate between our two simulation models is the transverse extent $\delta y$ of the source. We define $\delta y$ to be the full-width-half-maximum (FWHM) of the logarithm of the vertical intensity distribution $I_\nu(x_c, y)$ at the flux centroid coordinate $x_c$. Simulation measurements of the transverse extent are depicted as black bars in Figures \ref{fig:image-comparison-20} and \ref{fig:image-comparison-30}. In certain images, such as the 100 day image in the bottom of Figure \ref{fig:image-comparison-30}, $I_\nu(x_{\rm max}, y)$ crosses its half-maximum value twice on each side of the origin. In such cases we define $\delta y$ as the larger of the two possibilities. We have also plotted the time evolution of $\delta y$ for various observer angles and both explosion models in the bottom row of Figure \ref{fig:centroid-extent-comparison}. We find that, at times later than $\sim 100$ days, $\delta y$ is generally larger by $\simeq \unit[1]{mas}$ in the choked jet scenario than in the successful jet scenario.

\section{Summary}
\label{sec:summary}

We have computed synthetic radio image sky maps based on our previously published \citep{Xie2018} numerical simulation data of the two leading models for the explosion geometry of GRB170817A. We have done so to aid in the interpretation of possible late-time radio observations with sub-milliarcsecond resolution. Such observations would rule out a spherical explosion were they to show the flux centroid to be offset from the merger location. However, our results indicate a measurement of the flux centroid location may not easily differentiate between the successful and choked jet models. We find that all models predict that the centroid will exhibit superluminal mean proper motion of $\unit[4 - 10]{\mu as/day}$ for at least the first $\simeq$ 300 days. The absence of a dramatic difference is mainly due to the mildly relativistic expansion velocity of the explosion at times late enough that the blast might be spatially resolved.

Our analysis nevertheless reveals features, potentially discernible from well resolved, highly sensitive, and appropriately timed radio observations, that could be probes of the explosion geometry. We find that the time $t_{\rm max}$ when the centroid attains its maximum offset and begins moving back toward the merger site is potentially detectable and distinguishes between models. For the structured jet model $t_{\rm max}$ is a robust feature occurring at $t_{\rm max} \simeq 200$ days for $\theta_{\rm obs} = 20^\circ$ and $t_{\rm max} \simeq 400$ days for $\theta_{\rm obs} = 30^\circ$. The choked jet model, by contrast, plateaus near 600 days and likely lacks a detectable turnaround. Detection of a turnaround before 600 days would thus uniquely identify a successful jet.

Since this manuscript was first submitted, observations by \cite{Mooley2018} have been published indicating that the source was unresolved at 230 days. The FWHM values we reported in Section \ref{sec:results} are consistent with this measurement. However we also found that the choked jet model generally predicts a larger image than the successful jet. We predict that if VLBI observations were obtained at $\sim 400$ days, they would be marginally resolved for either model. A larger FWHM value of $\unit[7]{mas}$ favors a choked jet while a smaller value of $\unit[4.5]{mas}$ favors a successful jet.

\acknowledgments The authors acknowledge Brian Metzger and Sjoert van Velzen for valuable discussions.

\bibliographystyle{apj}

\end{document}